
\voffset=-1 cm
\font\twelvebf=cmbx10 scaled \magstep1
\hsize 15.2true cm
\vsize 22.0true cm
\voffset 1.5cm
\nopagenumbers
\headline={\ifnum \pageno=1 \hfil \else\hss\tenrm\folio\hss\fi}
\pageno=1

\hfill IUHET--398

\vskip 1cm

 
\noindent{\twelvebf SPIN-DEPENDENT FORCES BETWEEN QUARKS
IN HADRONS}
 
\bigskip
\noindent  D. B. Lichtenberg$^a$ 

\noindent Physics Department, Indiana University,
Bloomington, IN 47405, USA
\bigskip
\hskip 2cm (Invited talk to be given at Orbis Scientiae, 
Fort Lauderdale, 

\hskip 2cm December 18--21, 1998. A revised version will be
published

\hskip 2cm in the proceedings.)

 \vskip 1 cm

\item{} {\bf Abstract.} Different mechanisms have
been proposed to account for the spin-dependent
interaction between quarks in ground-state hadrons. The 
first mechanism is the chromomagnetic interaction
arising from one-gluon exchange, a second mechanism 
is an interaction arising from meson exchange, and 
a third from instanton effects.  Some
lattice calculations favor the first mechanism for
heavy quarks and the third for light quarks. 
An argument is presented in favor of the one-gluon
exchange mechanism. However,
thus far, nobody has performed a crucial experiment that
can definitively distinguish between these mechanisms.

\vskip 1cm

$^a$lichten@indiana.edu

\vskip 2cm


Most of what I say in this talk is within the framework
of the constituent quark model, in which explicit
gluon degrees of freedom are integrated out, leaving
only a potential between quarks. In this 
framework, the
question, ``What per cent of the spin of a hadron
is carried by the quarks?''  has the simple answer,
``100 per cent.'' The sea of gluons and quark-antiquark 
pairs surrounding a current quark is considered an integral
part of it, and both the quark and the sea together make 
the constituent quark. When a constituent quark is tweaked
gently, it drags its sea around with it, and therefore
has more inertia than a current quark. 
The size of a constituent quark,
including its sea, is substantially greater
than the size of a current quark, and 
a constituent quark may not be much smaller than 
the hadron in which it is bound.

In the constituent quark model there is still 
a puzzle about the nature of the spin-dependent
force between two quarks (or between a quark and an
antiquark). Specifically, does the spin-dependent
interaction arise from the chromomagnetic
interaction of one-gluon exchange [1], from 
Goldstone-boson exchange in connection with the
breaking of chiral symmetry [2], or from the effects of
instantons [3]? Here, arguments are given that the
one-gluon-exchange mechanism is the dominant one. 
Isgur [4] has also taken this point of view.

Shortly after the invention of QCD, De R\'ujula et al.\
[1] discussed consequences of the one-gluon exchange
interaction between two  (constituent) quarks in a
baryon and between a quark and antiquark in a meson. 
In their work, the spin-dependent splitting of ground-state
hadrons arises from the chromomagnetic interaction. 

Since then, many authors have written papers in which
the mass splittings of the vector and pseudoscalar 
mesons, and the splittings of the spin-3/2 and spin-1/2
baryons arise from chromomagnetic forces. These
forces have three salient features: (1) they are two-body
forces, (2) they are short range, and (3) they are
attractive in spin-0 states and repulsive in spin-1
states of two quarks in a baryon or a quark-antiquark 
in a meson.

Although the chromomagnetic interaction is a two-body
operator, Cohen and Lipkin [5] and Richard et al.\ [6]
pointed out that in baryons (and
more generally in hadrons containing more than two
quarks) the spin-dependent interaction {\it energy} 
between two quarks is influenced by the remaining
``spectator'' quarks. Anselmino et al.\ [7] gave
a quantitative measure of this effect in baryons. In 
cases in which data were available, they found that the
spectator quark influenced the interaction energy by
less than 12 \%. 

More recently, it was proposed that the spin-dependent
interaction between light quarks arises, not from one-gluon
exchange, but from Goldstone-boson exchange, the boson
being a pseudoscalar meson connected with chiral symmetry 
breaking. I call this the one-meson-exchange model;
see the  review by Glozman  and  Riska [2] for references. 
A third suggestion is that
instantons give the dominant contribution to the 
the spin-dependent interaction in light quarks. A
lattice gauge calculation [8] appears to support this
third point of view. I call this the instanton model;
references can be found in the review by Shuryak [3].

At present, all three models apparently have enough
flexibility to approximate the existing data. 
An important question, which has not yet been definitively
answered, is whether the three models are just different
ways of describing the same thing or whether their
predictions are different enough so that future
experiments may rule out at least two of them as giving
the dominant contribution to the spin splittings. 

Because of asymptotic freedom, there is every reason
to believe that at sufficiently small separation $r$ between 
quarks, QCD perturbation theory works well. Furthermore,
the chromomagnetic interaction of one-gluon exchange 
has a  short range, and so is more amenable to perturbative
treatment than the forces of longer range. Also, QCD
dynamics causes hadrons
containing only heavy quarks to be smaller in size than
hadrons containing only light quarks or both light and
heavy quarks. For threse three reasons, 
one-gluon exchange is expected to be the dominant
interaction at small distances in heavy-quark hadrons;
in particular, the chromomagnetic interaction
is expected to be the dominant mechanism of spin splittings
in such hadrons.

In my opinion the one-gluon exchange force also provides
a useful description of the spin-dependent splittings
in hadrons containing light quarks. My argument is
that the one-gluon-exchange model leads to qualitative
agreement with the fact that spin splittings are larger
in hadrons containing light quarks than in hadrons
containing only heavy quarks. I shall present this
argument in some detail. 

The spin-spin term $H_s$ of the one-gluon-exchange 
interaction has the form 
$$H_s = \nabla ^2 V \sigma_1 \cdot \sigma_2/(6 m_1 m_2),
\eqno(1) $$
where $V$ is the color-Coulomb potential,
$\sigma_1$ and $\sigma_2$ are Pauli spin matrices, and
$m_1$ and $m_2$ are the constituent quark masses. 
The quantity $V$ is given by 
$$V= \lambda_1 \cdot \lambda_2 \alpha_s /(4r), \eqno(2)$$
where $\lambda_1$ and $\lambda_2$ are Gell-Mann SU(3)
matrices, $\alpha_s$ is the strong coupling constant,
and $r$ is the distance between quarks.

The form given in Eq.\ (1) holds for a more general 
potential than the one-gluon exchange potential of
Eq.\ (2), as has been noted by a number of authors.
References can be found in a review of the potential
model [9]. The result is that
if the quark-quark (or quark-antiquark) static
potential $U$ is written as
$U=V + S$, where $V$ transforms as the time component
of a vector and $S$ transforms as a scalar, then it
is just the vector part $V$ that enters Eq.\ (1). It is
commonly assumed [9] that the short-range part of the 
potential is $V$, whereas the confining part is $S$. 

The interaction $H_s$ is repulsive if two quarks in
a baryon or a quark and antiquark in a meson have spin
1, and is attractive if the two quarks have spin 0. 
In the one-gluon-exchange model, this result is responsible
for the spin splittings in ground-state mesons and baryons.

I now examine how the expectation value 
$\langle H_s\rangle$ varies with the quark masses.
The factor $(m_1 m_2)^{-1}$ in Eq.\ (1) decreases for 
increasing 
quark masses, but this decrease might not be enough
to compensate for the fact that the expectation value
$\langle \nabla^2 V \rangle $ increases as the quark
masses increase. However, if the scalar confinement
term $S$ is linear and the $V$ of Eq.\ (2) is
treated as a perturbation, then Frank and O'Donnell [10]
have shown that  
$$\langle \nabla^2 V \rangle \propto \mu, \eqno(3)$$ 
where $\mu$ is the reduced mass of the two quarks.
It follows from Eqs.\  (1) and (3) that 
$$\langle H_s\rangle \propto (m_1 + m_2)^{-1}, \eqno(4)$$ 
a result that shows that indeed the effect of $H_s$
is larger in hadrons containing light quarks than in
hadrons containing heavy ones. The result (4) has
been generalized [11] to the case in which 
$$V \propto - r^{-n}, \quad S \propto r^n, \quad
0 < n \leq 1. \eqno(5)$$

Therefore, the one-gluon-exchange 
interaction of Eq.\ (1) accounts in a natural
way for the fact that spin splitting in light hadrons 
is larger than the splittings in heavy hadrons.
If the splitting in light
hadrons is to be explained by either one-meson exchange
or instantons, a way has to be found to ``turn off''
the one-gluon exchange contribution in light-quark
hadrons. I have not seen
any convincing way to do this. I must point out that 
Negele [8] claims that a lattice calculation shows
that the one-gluon-exchange contribution is negligible in
light mesons, but how that happens is mysterious to me.

According to the one-gluon-exchange model with a 
quark-antiquark potential of Eq.\ (5), the 
splitting $\Delta_M$ between the masses of 
ground-state vector and pseudoscalar mesons should
go like Eq.\ (4). To check this against experiment,
one needs a set of constituent quark masses. For 
definiteness, I take quark masses given by 
Roncaglia et al.\ [12]. (These quark masses lead to a hadron
binding energy that varies smoothly with the reduced 
mass of the quarks.) The quark masses are (in MeV)
$$m_u= m_d = 300, \quad m_s = 475, \quad
m_c = 1640, \quad m_b = 4985. \eqno(6) $$

In Table 1 are given the experimental mass splittings 
$\Delta_M$ from the Particle Data Group [13] and the values
of $R\Delta_M$, where $R = (m_1 + m_2)/(2m_u)$. 
If the potential is of the form given in Eq.\ (5) and
if the short-range part of the potential can be treated
as a perturbation, then
the values of $R \Delta_M$ should be the same constant
for all the ground-state splittings. 

\bigskip

TABLE 1. Values of the mass differences $\Delta_M $
between vector
and pseudoscalar mesons containing the same quark content.
Also given are the values of $R \Delta_M$, where $R =
(m_1 + m_2)/(2m_u)$, with the values of the $m_i$ 
given in Eq.\ (6).  Values of $\Delta_M$
are from the Particle Data Group [13]. According to the
one-gluon-exchange model, the values of $R\Delta_M$
should be approximately constant. 
 
\vskip 12pt
$$\vbox {\halign {\hfil #\hfil &&\quad \hfil #\hfil \cr
\cr \noalign{\hrule}
\cr \noalign{\hrule}
\cr
Mesons & Quark content$^{\dagger}$ & $\Delta_M$ (MeV) & 
$R\Delta_M$ (MeV) \cr
\cr \noalign{\hrule}
\cr
$\rho - \pi$ & $q \bar q$ & 632 & 632 \cr
$K^* - K$ & $q \bar s$ & 398 & 514 \cr
$D^* - D$ & $c \bar q$ & 141 & 456 \cr
$D^*_s-D_s$ & $c \bar s$ & 144 & 508 \cr
$J/\psi -\eta_c$ & $c \bar c$ & 117 & 640 \cr
$B^* - B$ & $q \bar b$ & 46  & 405 \cr
$B_s^* -B_s$ & $s \bar b$ &  47 & 428 \cr
 
\cr \noalign{\hrule}
\cr \noalign{\hrule}
}}$$

$^{\dagger}$The symbol $q$ means either $u$ or $d$.

\bigskip

It can be seen from Table 1 that the values of 
$R\Delta_M$ are not constant, but they are much 
closer to a constant than the values of $\Delta_M$. 
In my opinion, the fact that the values of $R \Delta_M$
do not vary by much (the smallest value is 63\% of 
the largest)
is a reflection of the underlying validity of the
one-gluon-exchange model. If so, the variations
in $R\Delta_M$ just show that using a potential
in a nonrelativistic Schr\"odinger equation and 
treating the short-range part of the potential as a 
perturbation are not adequate  approximations.
Note that the values of $R\Delta_M$ are largest
for the $\rho-\pi$ and $J/\psi-\eta_c$ splittings. 
In the case of the pion, the short-range 
spin-dependent force is sufficiently strong
to increase the pion wave function near $r= 0$, thereby
decreasing the pion mass more than one would predict from
perturbation theory. In the case of the $c\bar c$ system,
the color-Coulomb force increases the wave function
near $r=0$, again enhancing the effect of the short-range
spin-dependent force.

The effects of the chromomagnetic interaction
in ground-state baryons 
is a little more complicated than in mesons because a 
baryon contains three quarks. However,  the effect of the
two-body spin-dependent force can be approximately
isolated [7]. I distinguish two cases: first, the
case in which the baryons have two identical quarks,
and second, the case in which all  three quarks have
different flavors. If two quarks have the same flavor,
I define the spin-dependent splitting $\Delta_B$
by $\Delta_B = B^* - B'$, where $B^*$ is the
mass of the spin-3/2 baryon and $B'$ is the mass of
the spin-1/2 baryon. For example, in the case of baryons
made of only $u$ and $d$ quarks, the $B^* $ is the
$\Delta$ baryon and the $B'$ is the nucleon. In the
case in which all three quarks are different, there
exists a second spin-1/2 baryon, which I denote by $B$.
The $B'$ has the spin structure of the $\Sigma$ baryon,
while the $B$ has the spin structure of the $\Lambda$.
If all quarks are different, I define another
independent spin splitting $\Delta'_B =
(2B^* + B' -3 B)/2$. The quantity $R_B$ is defined
similarly to $R$ for mesons, except that the quark
masses are those of the active quarks. The two active
quarks are those responsible for the spin-dependent
force given in Eq.\ (1); the third quark is the 
spectator quark. Of course, to obtain the spin
splittings in baryons, one must appropriately sum
over all three pairs of quarks. 

I do not have
a derivation that $R_B\Delta_B$ or $R_B\Delta_B'$ should
be constant, but consider these quantities in analogy with 
the mesons.  In Table 2 are given the experimental values 
of spin splittings in ground-state baryons from the 
Particle Data Group [13]. Also given are the two active
quarks and the spectator quark (which has a small
effect on the spin splitting caused mainly by the
active quarks [7]).  

\bigskip

TABLE 2. Values of the mass differences $\Delta_B $
and $\Delta'_B$
between spin-3/2 and spin-1/2 baryons
containing the same quark content. See the text
for the definitions of the quantities. The table
distinguishes between the active quarks, which are mainly
responsible for the spin splittings,  and the spectator 
quark, which plays only a minor role.
Also given are values of $R_B \Delta_B$, where $R_B =
(m_1 + m_2)/(2m_u)$, with the values of the masses
of the active quarks $m_1$ and $m_2$
given in Eq.\ (6).  Values of $\Delta_B$ and $\Delta_B'$
are from the Particle Data Group [13].

\vskip 12pt
$$\vbox {\halign {\hfil #\hfil &&\quad \hfil #\hfil \cr
\cr \noalign{\hrule}
\cr \noalign{\hrule}
\cr
Baryons & Active  & Spectator & $\Delta_B$  &
$R_B\Delta_B$ \cr
& quarks$^{\dagger}$ & quark & or $\Delta_B'$ (MeV)
& or $R_B\Delta_B'$ (MeV)
\cr \noalign{\hrule}
\cr
$\Delta - N$ & $q  q$ & $q$ & 293 & 293 \cr
$\Sigma^* - \Sigma$ & $ qs$ & $q$ & 192 & 248 \cr
$\Xi^* - \Xi$ & $q  s$ & $s$ & 215 & 278 \cr
$\Sigma_c^* - \Sigma_c$ & $ qc$ & $q$ & 65  & 210 \cr
$(\Sigma^* + \Sigma- 3\Lambda)/2$ & $ qq$ & $s$ & 
307 & 307 \cr
$(\Sigma_c^* + \Sigma_c- 3\Lambda_c)/2$ & $ qq$ & $c$ 
& 317  & 317 \cr
\cr \noalign{\hrule}
\cr \noalign{\hrule}
}}$$
 
$^{\dagger}$The symbol $q$ means either $u$ or $d$.

\bigskip

It can be seen from the first, second, and fouth entries 
of the last column of Table 2 that in baryons the
spin-dependent splittings decrease a little faster
than $(m_1 + m_2)^{-1}$. In column 4,
it can be seen from a comparison of  the first, fifth,
and sixth entries and from a comparison of
the second and third entries that the
heavier the spectator quark, the larger the matrix 
element of the two active quarks. This is in agreement
with the conclusion of [7], and is the result of the
heavy spectator quark shrinking the hadron wave function
and thereby making the short-range spin-dependent 
interaction more effective.  

I have given a plausibility argument that the
one-gluon-exchange model is relevant for hadrons
containing light, as well as heavy, quarks. 
It is important to note that
nothing I have said rules out the possibility
that the one-meson-exchange model and the instanton
model also contribute in some measure.
For example, Buchmann et al.\ [14] conclude that
the $\Delta -N$ mass splitting arises 2/3 from
one-gluon exchange and 1/3 from one-meson exchange.

It would be desirable to let experiment decide
which of the three models is dominant in light and in
heavy hadrons. Buchmann [15] has shown that 
one-gluon exchange, but not one-meson exchange,
contributes to the quadrupole moment of the $\Omega^-$
baryon, but this moment has not been measured. Also,
an existing calculation shows that
the one-gluon-exchange and one-meson-exchange models 
give different predictions 
for the properties of some exotic hadrons [16]. However,
at the present time, too few exotic hadrons have been
identified to distinguish between models. 
Further progress will require additional  
theoretical and experimental work. 


\bigskip

I have benefited from discussions with
Alfons Buchmann and Floarea Stancu, but they do not 
necessarily agree with the conclusions of this paper. 
This work was supported in part by the
U.S. Department of Energy.

\bigskip 


\vskip 1cm
\noindent {\bf REFERENCES}
\medskip
\font \eightrm=cmr9
\par
\everypar{\parindent=0pt \hangafter=1\hangindent=2 pc}

\eightrm
\par
\noindent 1. A. De R\'ujula, H. Georgi, and S. L. Glashow,
{\it Phys.\ Rev.\ D} 12:147 (1975).

2. L.Ya Glozman  and D.O. Riska, {\it Phys.\ Rep.}\
268:264 (1996).

3. E.V. Shuryak, {\it Rev.\ Mod.\ Phys.}\ 65:1 (1993).  

4. N. Isgur, at Baryons '98 international conference, Bonn, 
Sept. 22-26, 1998. 

5. I. Cohen, H.J. Lipkin, {\it Phys.\ Lett.}\ 106B:119
(1981).

6. J.M. Richard, P. Taxil, {\it Ann.\ Phys.\ (N.Y.)} 
150:26 (1983).

7. M. Anselmino, D.B. Lichtenberg, and E. Predazzi,
{\it Z.\ Phys.\ C} 48:605 (1990).

8. J.W. Negele, in {\it Intersections between Particle and
Nuclear Physics}, Big Sky, Montana, May, 1997, AIP 
Conference Proceedings 412, edited by T.W. Donnelly,
AIP, Woodbury, New York (1997), p. 3.

9. D.B. Lichtenberg, {\it Intl.\ J. Mod.\ Phys.\ A} 2:1669
(1987).

10. M. Frank and P.J. O'Donnell, {\it Phys.\ Lett.\ B} 
159:174 (1985).

11. D.B. Lichtenberg, {\it Phys.\ Lett.\ B} 193:95 (1987).

12. R. Roncaglia, D.B. Lichtenberg, and E. Predazzi,
{\it Phys.\ Rev.\ D} 52:1722 (1995).

13. Particle Data Group: C. Caso et al., {\it Euro.\ Phys.\
J. C} 3:1 (1998).

14. A.J. Buchmann, E. Hern\'andez, and E. Faessler,
{\it Phys.\ Rev.\ C} 55:448 (1997).

15. A.J. Buchmann, {\it Z. Naturforsch} 52a:877 (1997).

16. Fl.\ Stancu, {\it Few Body Phys.\ Suppl.}, to be 
published (Proc. of the 
Workshop on $N^*$ Physics and Non-perturbative
QCD, Trento, Italy, May 18-29, 1998).

\bye